\documentclass[reprint,superscriptaddress,aps,prx,nofootinbib]{revtex4-2}

\usepackage{amsmath}
\usepackage{amssymb}

\usepackage{subfigure}

\usepackage{amsxtra}
\usepackage{mathrsfs}
\usepackage{amsfonts}
\usepackage{arydshln}
\usepackage{dsfont}
\usepackage{dcolumn}
\usepackage{bm}
\usepackage{graphicx}
\usepackage{epstopdf}
\usepackage{txfonts}
\usepackage{braket}
\usepackage{comment}
\usepackage{blkarray}
\usepackage{xcolor}
\usepackage{color, colortbl}

\usepackage{tikz}
\usepackage{tcolorbox}

\usepackage{color}
\usepackage{subfigure}

\usepackage{booktabs} 
\usepackage{siunitx} 
\usepackage{threeparttable} 
\usepackage{tabularx}
\usepackage{multirow}
\usepackage{diagbox}

\definecolor{niceblue}{rgb}{0.388235, 0.627451, 0.847059}
\definecolor{nicered}{rgb}{0.7,0.1,0.1}
\definecolor{nicegreen}{rgb}{0.1,0.5,0.1}
\definecolor{darkmagenta}{rgb}{0.55, 0, 0.55} 
\definecolor{persianblue}{rgb}{0.11, 0.22, 0.73}
\definecolor{LightCyan}{rgb}{0.88,1,1}

\usepackage[
	colorlinks=true,
	citecolor=green!50!black,
        linkcolor=persianblue!70!black,
	urlcolor=green!50!black,
	hypertexnames=false]{hyperref}

\usepackage{orcidlink}

\usepackage{atbegshi,picture}
\AtBeginShipoutNext{\AtBeginShipoutUpperLeft{%
  \put(\dimexpr\paperwidth-1cm\relax,-1.5cm){\makebox[0pt][r]{\footnotesize \texttt{FERMILAB-PUB-25-0050-T, UCI-HEP-TR-2025-03, IFT-UAM/CSIC-25-13}}}%
}}

\graphicspath{{figures/}{figures_appendix/}}

\begin{document}

\title{
Clash of the Titans: ultra-high energy KM3NeT event versus IceCube data
}

\newcommand{\fnal}{\affiliation{Theoretical Physics Department, Fermilab, P.O. Box 500, Batavia, IL 60510, USA}}

\author{Shirley Weishi Li\,\orcidlink{0000-0002-2157-8982}}
\email{shirley.li@uci.edu}
\affiliation{Department of Physics and Astronomy, University of California, Irvine, CA 92697}

\author{Pedro~Machado\,\orcidlink{0000-0002-9118-7354}}
\email{pmachado@fnal.gov}
\fnal

\author{Daniel Naredo-Tuero\,\orcidlink{0000-0002-5161-5895}}
\email{daniel.naredo@ift.csic.es}
\affiliation{Departamento de F\'{\i}sica Te\'orica and Instituto de F\'{\i}sica Te\'orica UAM/CSIC,\\
Universidad Aut\'onoma de Madrid, Cantoblanco, 28049 Madrid, Spain}

\author{Thomas Schwemberger \orcidlink{0000-0002-8471-6879}}
\email{tschwem2@uoregon.edu}
\affiliation{Department of Physics and Institute for Fundamental Science, University of Oregon\\
Eugene, Oregon 97403, USA}

\date{\today}

\begin{abstract}
KM3NeT has reported the detection of a remarkably high-energy through-going muon.
Lighting up about a third of the detector, this muon likely originated from a neutrino exceeding 10~PeV in energy.
The crucial question we need to answer is where this event comes from and what its source is.
Intriguingly, IceCube has been operating with a much larger effective area for a considerably longer time,  yet it has not reported neutrinos above 10~PeV.
We quantify the tension between the KM3NeT event and the absence of similar high-energy events in IceCube under various assumptions of the origin for the neutrino including the isotropic diffuse flux, cosmogenic flux, and a steady or transient point source. 
Through a detailed analysis, we determine the most likely neutrino energy to be in the range of 23 -- 2400~PeV, and find a tension between the experiments ranging from 3.6 to 3.1~$\sigma$ for diffuse origins and 2.9 to 2.0~$\sigma$ for point sources. The lack of observation of high-energy events in IceCube seriously challenges the explanation of this event coming from any known diffuse fluxes.
Our results indicate the KM3NeT event is likely the first observation of a new astrophysical source.
\end{abstract}

\maketitle

\section{Introduction}
\label{sec:intro} 

High-energy astrophysical neutrinos provide a unique means to observe the universe at distances and energies not easily accessible through electromagnetic radiation.
Unlike photons or charged particles, neutrinos can travel vast distances across the universe without being absorbed or deflected, making them invaluable probes of astrophysical phenomena, particularly at identifying high-energy astrophysical sources.

The IceCube experiment has been a pioneer in the field of high-energy neutrino astronomy, observing neutrinos above the PeV scale. \cite{IceCube:2013cdw, IceCube:2013low} report the observation of neutrinos at $1.04\pm0.16$ and $1.14\pm0.17$~PeV. 
These observations have led to the discovery of an isotropic diffuse neutrino flux~\cite{IceCube:2020wum, IceCube:2020acn, IceCube:2021uhz, Silva:2023wol, IceCube:2024fxo}. 
Additionally, point sources such as blazars have been identified as potential sources of high-energy neutrinos~\cite{IceCube:2022der}.

Moreover, several large-scale neutrino observatories are under construction or in development, such as KM3NeT in the Mediterranean Sea~\cite{KM3NeT:2016zxf}, Baikal-GVD in Lake Baikal~\cite{Baikal-GVD:2018isr}, the Pacific Ocean Neutrino Experiment (P-ONE) in the Pacific Ocean~\cite{P-ONE:2020ljt}, and the planned IceCube-Gen2~\cite{IceCube-Gen2:2020qha}. 
The synergy among these detectors, which employ different detection techniques and cover complementary regions of the sky, will improve our ability to detect astrophysical neutrinos and identify sources~\cite{Aiello:2024jbp}.

Recently, the KM3NeT collaboration reported a remarkable event: a through-going muon that triggered 3,672 photomultiplier tubes (PMTs), roughly a third of the detector~\cite{KM3NeT_pres, KM3NeT:2025npi}. 
The collaboration inferred the energy of the neutrino that produced this muon to be between 72 -- 2600~PeV, making it the highest-energy neutrino event ever confirmed.

Although this observation is extraordinary, it raises important questions, such as the origin of this neutrino and its production mechanism~\cite{KM3NeT:2025aps,KM3NeT:2025bxl,Muzio:2025gbr,neronov2025,Fang:2025nzg,Dzhatdoev:2025sdi}.
Given IceCube's longer exposure time and larger effective area, one would expect it to have detected similar high-energy muons, which would help identify the source. 
However, no events exceeding 10~PeV have been reported by IceCube. 
This suggests a possible tension between the KM3NeT single-event observation and IceCube's lack of observation in more than ten years of data~\cite{KM3NeT:2025ccp}.

In this paper, we aim to examine potential sources for this KM3NeT event through compatibility tests of KM3NeT and IceCube data. 
We consider the diffuse flux observed by IceCube, cosmogenic neutrinos, and potential transient point sources. 
We evaluate the likelihood of each source by computing the tension between the KM3NeT detection and IceCube's non-observation given each flux assumption. 
Our analysis complements KM3NeT's test of the compatibility between their event and data from IceCube and Auger~\cite{KM3NeT:2025ccp}. 
While our results are qualitatively similar to KM3NeT's, the differences in procedures lead to quantitatively distinct outcomes. 
We comment on these differences throughout this manuscript.

\section{The KM3NeT event}\label{sec:neutrino-energy}

Triggering a significant fraction of the detector's PMTs, the KM3NeT event is consistent with a through-going muon~\cite{KM3NeT:2025npi}. 
The KM3NeT detector is still being constructed, and this event occurred during the ARCA21 phase, where 21 strings have been deployed, with an exposure of 288 days.
The angular direction of the muon is well reconstructed, coming from 0.6$^\circ$ above the horizon.
KM3NeT has estimated the muon energy to be 120 PeV, and the parent neutrino energy to be within 72 -- 2600~PeV at 90\% confidence level (CL).
In addition, the collaboration reported the number of PMTs that were triggered by the through-going muon, $N_\text{hit} = 3672$, as well as Monte Carlo simulations of the number of PMTs triggered by 10, 100 and 1000~PeV through-going muons~\cite{KM3NeT:2025npi}. 

\begin{figure}[t]
    \centering
    \includegraphics[width=1.0\linewidth]{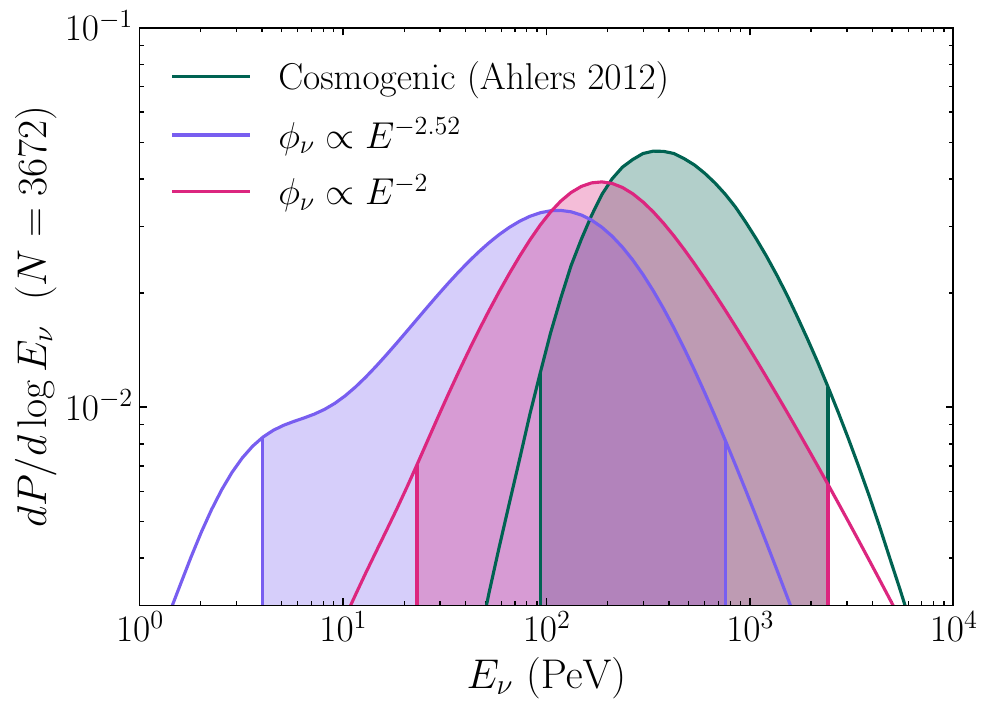}
    \caption{Inferred probability density of the neutrino energy of the KM3NeT event, given the 3,672 triggered PMTs, for three different underlying source flux assumptions.}
    \label{fig:nu_pdf}
\end{figure}

Even though the collaboration reported the range of likely neutrino energy for the event, it did not report its probability density, which our analysis requires. Therefore, we first infer the probability density for the neutrino energy. 
Bayes theorem states that the probability of a neutrino energy given the number of triggered PMTs (hits) is
\begin{equation}
    P(E_\nu|N_{\rm hit}) = \frac{1}{P(N_{\rm hit})} \int dE_\mu P(N_{\rm hit}|E_\mu)P(E_\mu|E_\nu)P(E_\nu),
\end{equation}
where $P(E_\nu)$ is the prior on the neutrino energy which we take to be proportional to the neutrino flux times effective area, namely $P(E_\nu)\propto d\phi/dE_\nu \cdot A_{\rm eff}(E_\nu)$.
{{$P(E_\mu|E_\nu)$ is the probability that an interaction of neutrino with energy $E_\nu$ outside the detector leads to a muon entering the detector with energy $E_\mu$.
We use \texttt{MadGraph}~\cite{Alwall:2014hca}} to obtain the kinematics of the hard scattering and we consider the average muon energy loss outside the detector volume.
The probability of a muon triggering $N_{\rm hit}$ PMTs, $P(N_{\rm hit}|E_\mu)$, depends on muon energy loss within the detector,} simulated here with \texttt{PROPOSAL}~\cite{koehne2013proposal}, as well as the details of the detector, for which we adopt a simplified model to account Cherenkov light production, light propagation in water, and light detection. 
In addition, we set a prior that muons entering the detector need at least 1~PeV energy.
For more details, see App.~\ref{app:Nhit}.

Figure~\ref{fig:nu_pdf} shows the inferred neutrino energy of the event
for three representative priors on the neutrino energy: 
a power-law flux $E_\nu^{-2.52}$,
which represents the diffuse flux observed by IceCube~\cite{Naab:2023xcz};
a power-law flux $E_\nu^{-2}$, which represents a generic source spectrum motivated by Fermi acceleration;
and the cosmogenic neutrino flux from Ahlers \emph{et al.}~\cite{Ahlers:2012rz}.
We also highlight the 90\% interval and most likely neutrino energies, summarized in Table~\ref{tab:intervals}.
We find excellent agreement when comparing our 90\% confidence interval (CI) with KM3NeT's official 90\% energy estimate for a $E^{-2}$ power-law~\cite{KM3NeT:2025npi}.
{We note the feature around 3 PeV in the $E^{-2.52}$ power-law case is due to the  prior on the neutrino spectrum which dominates over the  conditional probability $P(N_{\rm hit}|E_\nu)$ at low energies.}
While the 90\% interval has some dependence on the flux prior, we can see a robust preference for neutrino energies at or above the 100 PeV scale, possibly reaching EeV energies.
This preference for ultrahigh neutrino energies raises an obvious question: 
Where does this event come from? Why hasn't IceCube observed similarly high-energy events?

\begin{table}[t]
    \centering
    \begin{tabular}{c|c|c}
        Prior & \hspace{1mm}Peak energy (PeV)\hspace{1mm} & 90\% CI (PeV) \\ \hline
        Power-law, $(\gamma=2.52)$ & 120 & [4.0,\hspace{2mm}760] \\
        Power-law, $(\gamma=2.0)$  & 190  & [23,\hspace{2mm}2400] \\
        Cosmogenic (Ahlers 2012)   & 335  & [93,\hspace{2mm}2400] \\
    \end{tabular}
    \caption{Most likely neutrino energies and 90\% confidence intervals for the KM3NeT event assuming different underlying fluxes.}
    \label{tab:intervals}
\end{table}

\section{Compatibility between KM3NeT and IceCube data}

The estimated compatibility between KM3NeT and IceCube data depends on the assumptions we adopt for the origin of the event.
We will investigate three possibilities for the origin of this event: diffuse power-law, cosmogenic, and point source neutrino fluxes.
For each assumption, the expected number of events triggering at least $N_{\rm hit}$ PMTs at KM3NeT can be calculated as
\begin{equation}\label{eq:N_expected}
    N_{\rm evt}=4\pi T \!\!\int_0^\infty\!\!\! dE_\nu\!\int_{N_{\rm hit}}^{\infty}\!\!  d\hat N_{\rm hit}\, P(\hat N_{\rm hit}|E_\nu) \,\left\langle\frac{d\phi}{dE_\nu} A_{\text{eff}}(E_\nu)\right\rangle_\Omega\,,
\end{equation}
where $\phi$ is the neutrino flux, $A_{\rm eff}$ is the effective area, angle brackets indicate an average over the solid angle, $T=288$~days~\cite{KM3NeT:2025npi} is the exposure, and $P(N_{\rm hit}|E_\nu) = \int dE_\mu P(N_{\rm hit}|E_\mu)P(E_\mu|E_\nu)$ is the probability that $N_{\rm hit}$ PMTs are triggered by a muon originating from a neutrino with energy $E_\nu$, as discussed in the previous section.

\subsection{Power-law diffuse flux}

The diffuse flux of one flavor (muon neutrinos and antineutrinos) is well-characterized by a single power-law spectrum:
\begin{equation}
    \frac{d\phi}{dE_\nu} = \phi_0\left(\dfrac{E_\nu}{100 \, \mathrm{TeV}}\right)^{-\gamma} \times 10^{-18} / \text{GeV}/\text{cm}^2 / \text{s}/ \text{sr}\,,
    \label{eq:SPL}
\end{equation}
where $\phi_0$ is a dimensionless flux normalization and $\gamma$ is the spectral index.
This parametrization has successfully described the observed neutrino spectrum from tens of TeV to the PeV scale in IceCube~\cite{Naab:2023xcz}.

In order to quantify the compatibility between KM3NeT and IceCube, we employ the likelihood
\begin{equation}
    \mathcal{L}(\phi_0,\gamma)=\mathcal{L}_{\rm flux}(\phi_0,\gamma)\cdot\mathcal{L}_{\rm evt}(N_{\rm evt}(\phi_0,\gamma))\,.
    \label{eq:diffuse_likelihood}
\end{equation}
We account for IceCube's measurement of the high-energy neutrino flux with a Gaussian prior $\mathcal{L}_{\rm flux}$ in $\phi_0$ and $\gamma$.
Although the flux parameters from IceCube differ slightly from different datasets, they are largely consistent with each other~\cite{Naab:2023xcz}. 
For our analysis, we take the results of the combined fit~\cite{Naab:2023xcz}, which yields $\phi_0=1.80^{+0.13}_{-0.16}$ and $\gamma=2.52\pm0.04$.
The probability of observing one event at KM3NeT with $N_{\rm hit}\geq3672$, given the flux parameters, is encoded in the Poissonian term $\mathcal{L}_{\rm evt}=N_{\rm evt}e^{-N_{\rm evt}}$, see Eq.~(\ref{eq:N_expected}). 
The posterior distribution is computed from the likelihood~\eqref{eq:diffuse_likelihood} assuming a flat prior on $N_{\rm evt}$. 
{We then quantify the tension by means of the usual Bayes factor between the KM3NeT observation and IceCube preference:
\begin{equation}
    \mathcal{B}=\int \mathcal{P}(\phi_0,\gamma)\,\rm{d}\phi_0\rm{d}\gamma =\int \mathcal{L}(\phi_0,\gamma)\,\rm{d}\phi_0\rm{d}\gamma\,,
\end{equation}
where $\mathcal{P}$ is the posterior distribution which, given our flat prior on $N_{\rm evt}$, is equal to the likelihood $\mathcal{L}$ in Eq.~\eqref{eq:diffuse_likelihood}. 
Notice the lack of the denominator, as IceCube's likelihood on the flux parameters, which constitutes the null hypothesis, is normalised.}
\begin{figure}
    \centering
    \includegraphics[width=1.0\linewidth]{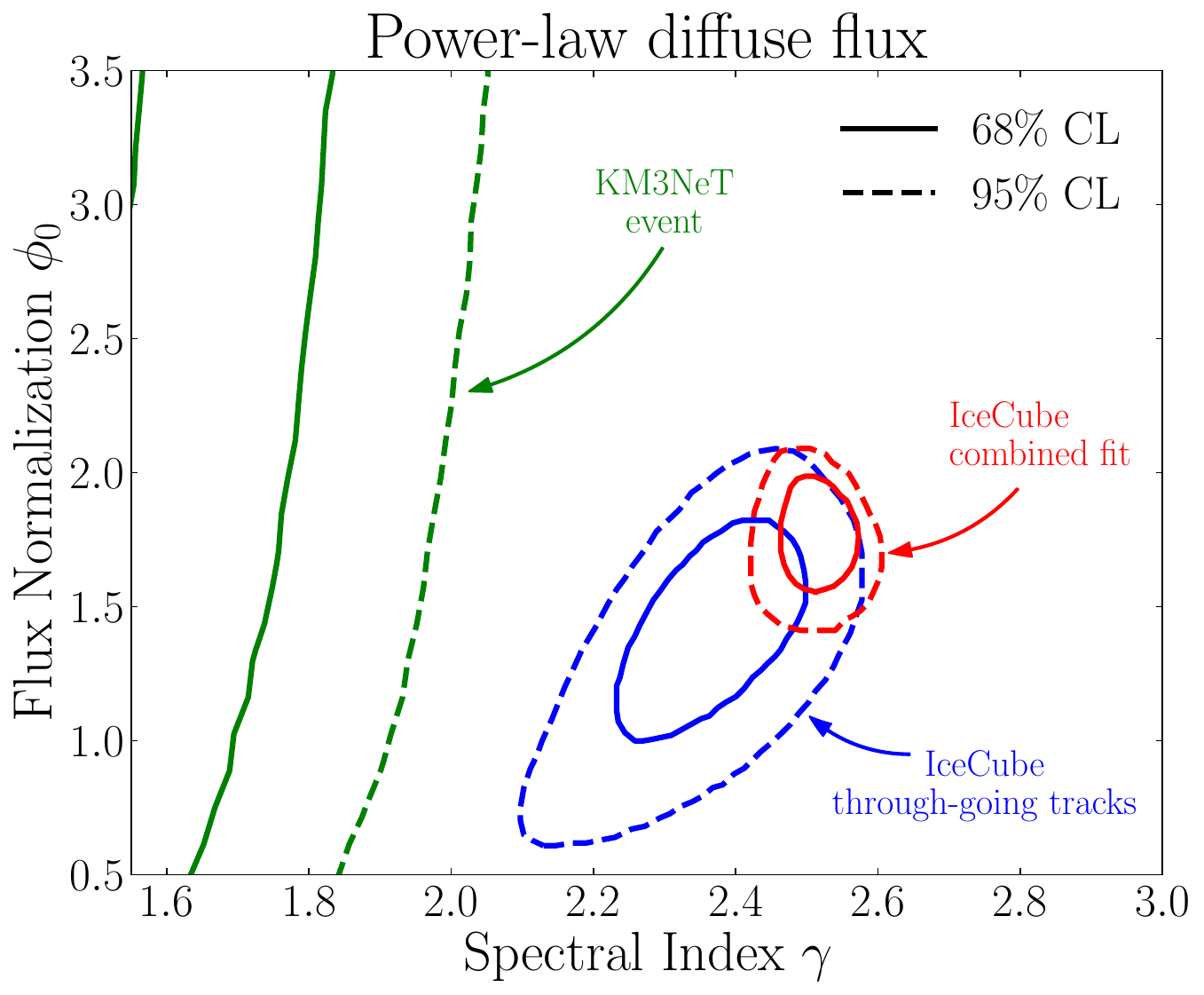}
    \caption{Preferred regions in the diffuse power-law spectrum flux normalization $\phi_0$ and spectral index $\gamma$ for the ultra-high energy KM3NeT event and IceCube data at 68\% and 95\% CL.}
    \label{fig:power-law-preferred}
\end{figure}

After marginalizing over the flux parameters, we find a $0.039\%~(3.5\sigma)$ tension between the ultra-high energy event observed in KM3NeT and the IceCube measurement of astrophysical neutrinos.
If we take IceCube's power-law preference based on through-going events only~\cite{IceCube:2021uhz}, which allows slightly harder spectra ($\phi=1.44^{+0.25}_{-0.26},~\gamma=2.37^{+0.09}_{-0.09}$), the tension is still $0.13\%~(3.2\sigma)$.

The tension found here is greater than the $1.9\sigma$ quoted by KM3NeT~\cite{KM3NeT:2025ccp}. The reason for this discrepancy is twofold. Firstly, KM3NeT calculates the tension assuming that the probability for the neutrino energy is flat in the 90\% CI. We account for the neutrino energy probability distribution, which penalizes lower energies where the tension would be smaller. Secondly, KM3NeT includes their non-observation of events outside the 90\% CL energy interval. 
This increases the degrees-of-freedom while barely contributing to the test-statistic, due to strong analysis cuts.
Consequently, this lowers the tension. 
If we follow KM3NeT's procedure, we confirm their $1.9\sigma$ result.

Another way to appreciate this tension is to examine the preferred regions for the power-law spectrum parameters given IceCube data and the ultra-high energy KM3NeT event in Fig.~\ref{fig:power-law-preferred}.
Since it is obtained by fitting a single event while dropping the Gaussian prior on Eq.~\eqref{eq:diffuse_likelihood}, the KM3NeT region exhibits a degeneracy between $\phi_0$ and $\gamma$.
For IceCube, we show the allowed regions for through-going events only and through-going plus shower events.
The lack of overlap between IceCube and KM3NeT is the tension among these data.
Given IceCube's nominal flux, KM3NeT should have expected 0.005 events between 72 -- 2600 PeV  during the exposure. 
Alternatively, given the normalization of the flux inferred from KM3NeT's event, assuming a spectral index of 2.52, IceCube should have expected 75 events in their diffuse search between 72 -- 2600 PeV.

\subsection{Cosmogenic flux}

Given the extremely high energy of the event, it is reasonable to ask if this is the first detection of the long-awaited cosmogenic neutrinos.
Cosmogenic neutrinos are produced when ultra high-energy cosmic rays interact with the cosmic microwave background~\cite{Greisen:1966jv, Zatsepin:1966jv}. 
These neutrinos, which can reach EeV energies, have never been observed.
We consider three benchmark scenarios for the cosmogenic flux, Ahlers 2010~\cite{Ahlers:2010fw}, Ahlers 2012~\cite{Ahlers:2012rz}, and van Vliet 2019~\cite{vanVliet:2019nse}, as possible origins of the KM3NeT event.
These scenarios have been previously studied by IceCube~\cite{IceCube:2025ezc}, which found $p$-values of 0.3\%, 4.3\%, and 26.8\% for these models respectively based on the non-observation of a neutrino flux exceeding $\sim100$ PeV. For each model, we calculate the probability (p-value) of producing an event in KM3NeT that triggers at least 3672 PMTs employing Poisson statistics. We combine this p-value with that of IceCube by converting each of them to a test-statistic value through a $\chi^2$-distribution with $1$ d.o.f. We then sum both values and convert back into a combined p-value by means of a $\chi^2$-distribution with $2$ d.o.f. The combined p-values are shown in Table~\ref{tab:cosmogenic_pvalues}.

\begin{figure}
    \centering
    \includegraphics[width=1.0\linewidth]{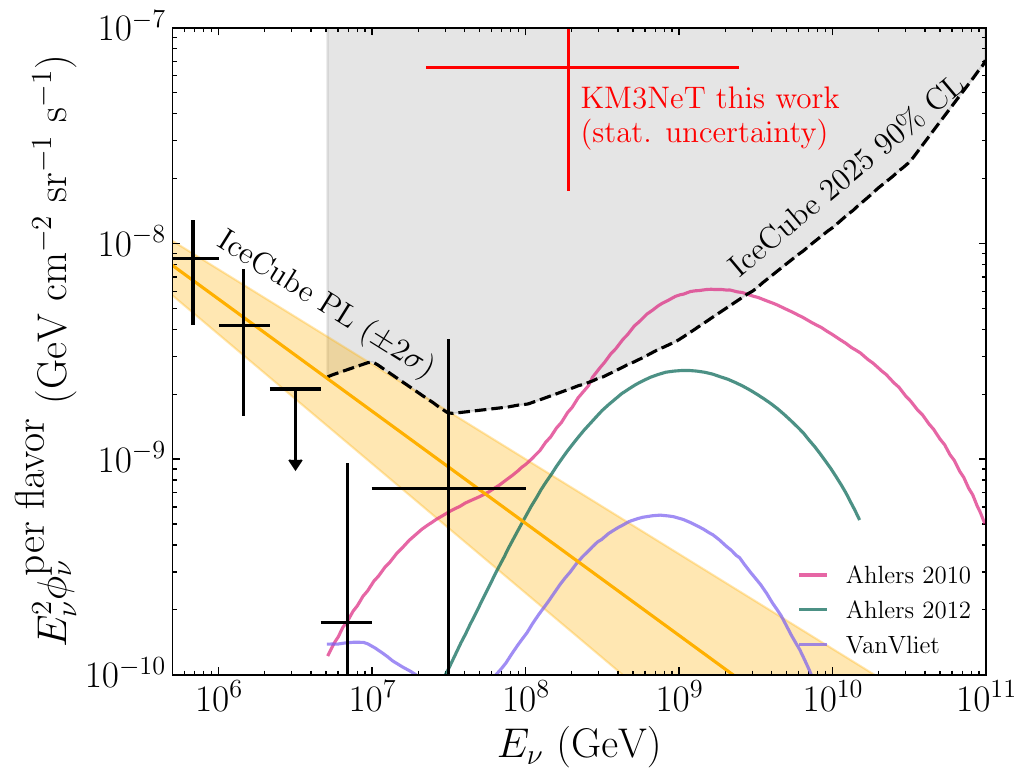}
    \caption{
    Predictions, measurements, and constraints on the high energy neutrino flux. The gold band shows the IceCube measurement of a single power-law flux, extrapolated to $E_\nu>100$~PeV while the three colored curves represent a selection of cosmogenic neutrino models~\cite{Ahlers:2010fw, Ahlers:2012rz, vanVliet:2019nse}. We show IceCube's segmented power-law data in black with error bars~\cite{Naab:2023xcz} and high energy constraints in gray~\cite{Meier:2024flg}. In red we show a best fit segmented power-law flux from the event in KM3NeT in the 90\% CI from a $E_\nu^{-2}$ power-law prior and $1\sigma$ error bars on the flux normalization. 
    }
    \label{fig:cosmogenic_fluxes}
\end{figure}

\begin{table}[t]
    \centering
    \begin{tabular}{c|c}
         \hspace{3mm}Cosmogenic Model \hspace{3mm}& \hspace{3mm}Combined $p$-value\hspace{3mm} \\ \hline
         Ahlers2010 1EeV& $3.3\times 10^{-4}$ ($3.6\sigma$) \\
        Ahlers 2012   & $1.6\times 10^{-3}$  ($3.1\sigma$)\\
        VanVliet2019  & $1.6\times 10^{-3}$  ($3.1\sigma$)\\
    \end{tabular}
    \caption{Combined $p$-value of KM3NeT and IceCube, for distinct cosmogenic neutrino flux models.}
    \label{tab:cosmogenic_pvalues}
\end{table}
We find $p$-values of $0.033\%$, $0.16\%$, $0.16 \%$ for Ahlers 2010, Ahlers 2012, and van Vliet 2019, respectively, revealing a robust tension between KM3NeT and IceCube data under these assumptions.
The reason for the tension is fairly simple and it is depicted in Fig.~\ref{fig:cosmogenic_fluxes}, where we show IceCube and KM3NeT  data, the flux predictions for each cosmogenic model, and the IceCube constraints on the power-law and cosmogenic fluxes. 
In a nutshell, Ahlers 2010 predicts $6.2\times10^{-3}$ events in KM3NeT but it is disfavored by IceCube with a $p$-value of 0.3\%.
Using Ahlers 2012 yields $2.8\times 10^{-3}$ events in KM3NeT but its IceCube $p$-value is 4.3\%.
Finally, van Vliet 2019, with the highest IceCube $p$-value of 26.8\%, only predicts $5.7\times 10^{-4}$ events in KM3NeT.
Therefore, none of the benchmarks can readily accommodate the observed ultra-high-energy event without being in tension with IceCube's non-observation of neutrinos above 10 PeV.
Even without taking IceCube data into account, none of the cosmogenic models considered here predict a sizable event rate in KM3NeT.


We note that the tension we report is more significant than KM3NeT’s official result. In Ref.~\cite{KM3NeT:2025vut}, they report a probability of 5.6\%, corresponding to $1.9\sigma$, for observing one neutrino event of cosmogenic origin during their exposure.
There are several differences between our analyses. We consider only four cosmogenic flux models that are most compatible with IceCube’s data, whereas KM3NeT evaluated broader classes of cosmogenic models while varying inputs such as the star formation rate, xx, and xx. Moreover, KM3NeT does not incorporate IceCube’s latest constraints on cosmogenic fluxes~\cite{IceCube:2025ezc}.


\subsection{Point Sources}

The tensions discussed in the previous sections could potentially be alleviated if the event seen in KM3NeT originated from either a steady point source or a transient source that was in an unfavorable location for IceCube detection.
At the high energies preferred by the KM3NeT event, absorption in the Earth strongly attenuates up-going neutrinos, making effective areas virtually null for sources below the detector horizon.
Similarly, for down-going events, the smaller amount of target material above the detector not only reduces the interaction probability but also increases the cosmic ray background, which suppresses the effective area.

\begin{figure}
    \centering
    \includegraphics[width=0.97\linewidth]{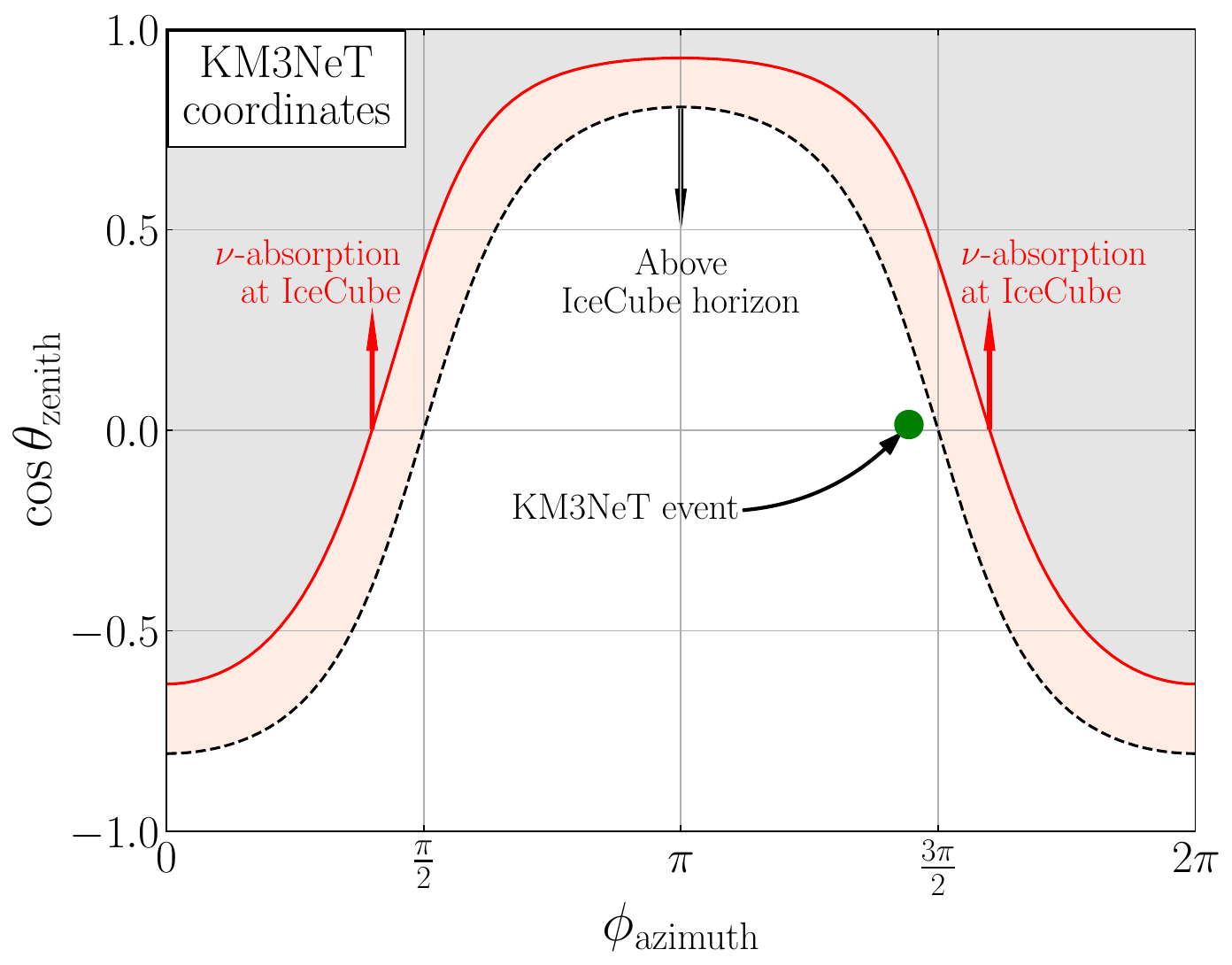}
    \caption{IceCube sky coverage on the KM3NeT local zenith and azimuth angles.}
    \label{fig:directionality}
\end{figure}

Therefore, a source location where KM3NeT has good sensitivity but IceCube has reduced sensitivity due to either Earth absorption or reduced interaction probability could help explain the apparent discrepancy between the detectors. 
For example, if the event originated from the north, it would be in the grey region of Fig.~\ref{fig:directionality}.
Nevertheless, this is not the case here.
Figure~\ref{fig:directionality} depicts the IceCube sky coverage on KM3NeT local zenith and azimuth coordinates.
The shaded gray region corresponds to a declination at IceCube of -15$^\circ$ where Earth matter would significantly absorb neutrinos above $E_\nu=10$ PeV~\cite{IceCube:2021xar}.
The dashed black line, on the other hand, corresponds to  the IceCube horizon.

The event in KM3NeT, shown as a green dot in Fig.~\ref{fig:directionality}, was within 1$^\circ$ of the local horizon providing maximal sensitivity for KM3NeT.
However, as shown in Fig.~\ref{fig:directionality} the event's azimuthal angle of approximately 259.8$^\circ$ at KM3NeT (36$^\circ$ latitude) corresponds to roughly 8$^\circ$ above the horizon at IceCube --where IceCube maintains optimal sensitivity to ultra-high energy neutrinos.

Since the event originated from a zenith angle where IceCube has a larger effective area and has been recording data for about 10 times longer, a point source is unable to alleviate the tension between the two experiments. 
For a single event in KM3NeT, one would expect the number of events in IceCube to simply scale up as the ratio of effective areas times exposures, namely,
$\left(A_{\rm eff}T\right)_{\rm IC} / \left(A_{\rm eff}T\right)_{\rm K}\sim 280$, where we have considered $10$ years of IceCube and averaged the effective areas over the 90$\%$ CL energy interval when assuming a $E_\nu^{-2}$ power-law prior (see Table~\ref{tab:intervals}). 
Notice that, due to the Earth's rotation, KM3NeT's effective area should be averaged over the $\cos\theta_{\rm zenith}$ range spanned in a day. 
We find that $\cos\theta_{\rm zenith}\in[-0.87,0.73]$, which implies that the day-averaged effective area is close to the sky-averaged effective area.\footnote{Note that KM3NeT's angular effective area is not public. We validate this assumption using the angular dependence of IceCube's effective area~\cite{IceCube:2021xar} finding that the daily and all sky averages at the KM3NeT location vary by $\sim 10\%$.}
Hence, we employ the latter in our statistical procedure.
We consider a likelihood consisting of two Poissonian terms, which encode the probability of observing one event at KM3NeT and none at IceCube when $N_{\rm K}$ and $N_{\rm IC}$ are expected respectively.
\begin{equation}
    \mathcal{L}=N_{\rm K}e^{-N_{\rm K}}\cdot e^{-N_{\rm IC}}\hspace{0.5cm}\text{with}\hspace{0.5cm}\frac{N_{\rm IC}}{N_{\rm K}}=\frac{\left(A_{\rm eff}T\right)_{\rm IC}}{ \left(A_{\rm eff}T\right)_{\rm K}}\equiv \rho\sim 280\,,
\end{equation}
{We test, by means of a Bayes factor, this hypothesis against the scenario in which no events are seen in both experiments, assuming flat priors on $N_{\rm K}$ and $N_{\rm IC}$, which can be computed analytically:
\begin{equation}
    \mathcal{B}=\dfrac{\int N_{\rm K}e^{-N_{\rm K}}e^{-\rho N_{\rm K}}\,\rm{d}N_{\rm K}}{\int e^{-N_{\rm K}}e^{-\rho N_{\rm K}} \,\rm{d}N_{\rm K}}=\dfrac{1}{1+\rho}\,.
\end{equation}
}
This boils down to a $2.9\sigma$ tension. 
Similar results are obtained using a frequentist approach.

A last possibility would be a transient source whose neutrinos only started arriving at the Earth after KM3NeT started taking data.
This would mitigate the tension by reducing IceCube's longer exposure.
Nevertheless, the difference in effective areas still lead to a $2.0\sigma$ tension.
We have considered the location of the event on the sky to be fixed at the best fit value from KM3NeT, though the collaboration reports a 90\% containment radius of $2.2^\circ$. At the limit of this radius, the event could be about $10^\circ$ above the horizon in IceCube which would slightly reduce these tensions, though the probability of such a scenario must also include the angular PDF of the source.
In that case, due to large cosmic muon backgrounds, it would not be surprising that IceCube missed this transient source.
However, as discussed in Ref.~\cite{KM3NeT:2025npi}, $5.6^\circ$ mis-reconstruction would be a $5\sigma$ deviation from KM3NeT's nominal systematic uncertainty on the direction of the event.

\section{Conclusions}

The extraordinary event observed by KM3NeT, triggering one-third of KM3NeT-ARCA21 photomultiplier tubes, begs the question of its origin.
We have shown that, if due to a neutrino interaction, this event would likely point to a neutrino with energies of hundreds of PeV.
While this would be the highest energy neutrino observed to date, the lack of such high-energy events in IceCube's data set seems to point to an important tension between these experiments.

We have analyzed this potential tension under several assumptions for the origin of the KM3NeT event: power-law diffuse flux; cosmogenic origins; and point sources.
Under any of these assumptions, the tension is found to be between $2.9\sigma$ and $3.6\sigma$, corresponding to a $p$-values of 0.36\% and $3.3\times 10^{-4}$, respectively.
Only a transient source would reduce this tension to $2.0\sigma$, or 5.5\%.
The robust tension between this event and data from IceCube deepens the mystery of its origin.
Given the small probability of this event to originate from the diffuse neutrino flux observed by IceCube, the KM3NeT event is very likely the first observation of a new astrophysical source.

\section*{Acknowledgements}
We thank John Beacom, Asher Berlin, Mattias Blennow, Mauricio Bustamante, Brian Clark, Enrique Fernandez-Martinez, Patrick Fox, Stefan Hoeche, Dan Hooper, Matheus Hostert, Joshua Isaacson, Tim Linden, Toni Mäkelä, Danny Marfatia, Nathan Whitehorn, and Bei Zhou for discussions.
This manuscript has been authored by Fermi Forward Discovery Group, LLC under Contract No. 89243024CSC000002 with the U.S. Department of Energy, Office of Science, Office of High Energy Physics.
The work of DNT is funded by the Spanish MIU through the National Program FPU (grant number FPU20/05333) and he has received support from the European Union’s Horizon 2020 research and innovation programme under the Marie Skłodowska-Curie grant agreement No~860881-HIDDeN and No 101086085 - ASYMMETRY, and from the Spanish Research Agency (Agencia Estatal de Investigaci\'on) through the Grant IFT Centro de Excelencia Severo Ochoa No CEX2020-001007-S and Grant PID2022
137127NB-I00 funded by MCIN/AEI/10.13039/501100011033.
TS is supported by the Dept. of Energy Office of Science Graduate Student Research (SCGSR) program administered by the Oak Ridge Institute for Science and Education (ORISE) for the DOE. ORISE is managed by ORAU under contract number DE-SC0014664.

\vspace{8mm}
\textbf{Note added:} {After the publication of the initial version of this manuscript, the official KM3NeT results came out~\cite{KM3NeT:2025npi}. 
We have updated all our inputs including the effective areas and the distributions of triggered PMTs accordingly.} 
Our conclusions remain unchanged.
Moreover, Ref.~\cite{neronov2025} appeared, claiming that for a transient point source, the consistency between IceCube and KM3NeT is below the 90\% CI.  
The main difference in our analyses is the statistical treatment in computing $p$-values.
Reference~\cite{neronov2025} computes the 90\% CI for both experimental observations, and due to the overlap of these two intervals, they conclude there is no tension at 90\% CI.
This procedure double counts the statistical error, artificially lowering the tension.

\appendix

\section{Effective area}\label{app:eff_area}

The sensitivity of neutrino telescopes is typically quantified by the effective area, $A_\text{eff}$, which includes the propagation and scattering of neutrinos in the Earth. In the following, we reproduce $A_\text{eff}$ from KM3NeT in order to extend it to higher neutrino energies.

By including propagation and scattering, the effective area captures all the ingredients to compute the neutrino event rate in an experiment.
To be more precise, the expected rate of events is given by
\begin{equation}
\frac{dN(E_\nu)}{dE} = T \int d\Omega \,A_\text{eff}(E_\nu, \cos\theta) \, \Phi(E_\nu, \Omega),
\end{equation}
where $\Omega$ is a solid angle, and $\Phi$ and $E_\nu$ are the neutrino flux and energy.
Following the prescription in Ref.~\cite{Gaisser:2016uoy}, we derive the effective area as
\begin{equation}
A_\text{eff}(E_\nu, \cos\theta) = \rho N_A V \sigma(E_\nu) e^{-\tau(E_\nu, \cos\theta)} \epsilon(E_\nu, \cos\theta),
\end{equation}
where $\rho$ is the density of the Earth near the detector, $N_A$ is Avogadro's number, $V$ is a volume, $\tau$ is the attenuation length of a neutrino with energy $E_\nu$, arriving from a zenith angle $\theta$, and $\epsilon$ captures the detector efficiency. 
We assume that for the high energies we are considering, the KM3NeT efficiency is unity, $\epsilon = 1$. 
We compute the attenuation of neutrinos in Earth from the neutrino cross section $\sigma(E_\nu)$ and using the PREM density profile~\cite{Dziewonski:1981xy}. 
For starting events, the volume $V$ is the same as the detector volume. 
However, for through-going muons, $V$ is more complicated as the effective volume is larger since muon neutrinos can interact outside the detector. 
This produces a muon which still propagates to the detector. 
We account for this effect following Ref.~\cite{Gaisser:2016uoy}. 
Although there is a dependence of the effective volume on the specific detector geometry, which translates into a $\cos\theta$ dependence to the effective area, this effect is much less significant than attenuation at PeV energies. 
For simplicity, we assume a spherical detector.
\begin{figure}[t]
    \centering
    \includegraphics[width=1.0\linewidth]{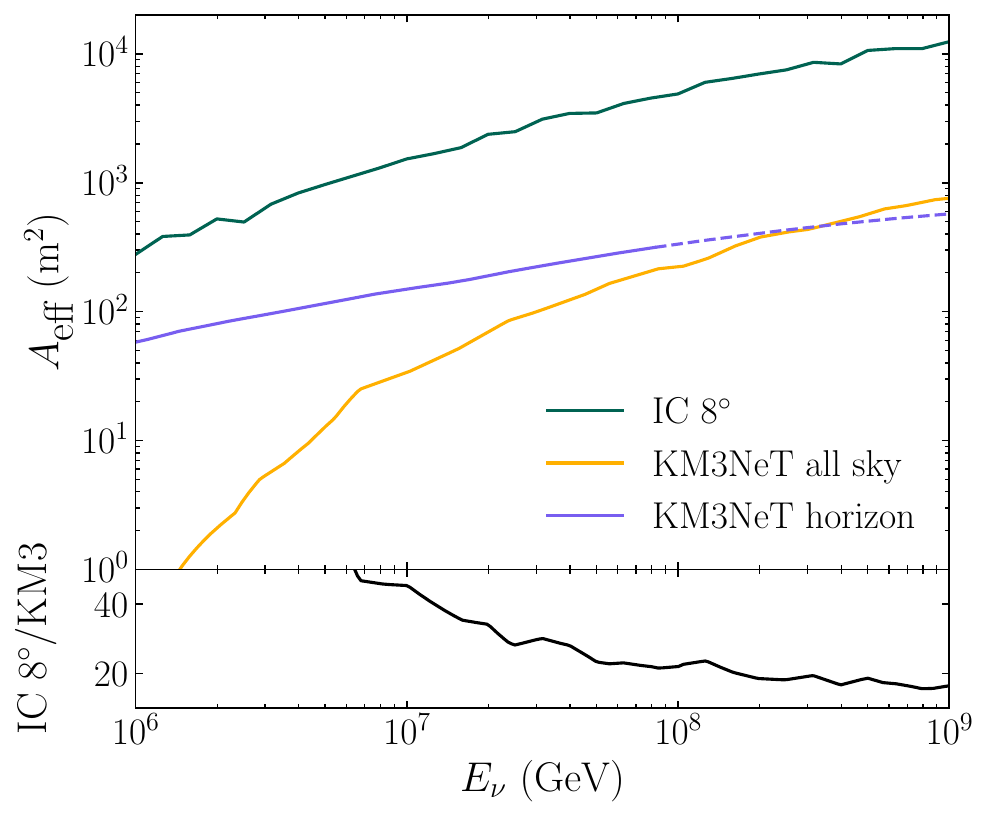}
    \caption{Effective areas used in this work. 
    We use the IceCube effective area $8^\circ$ above the horizon~\cite{IceCube:2021xar} (green) in our discussion of point sources. 
    For KM3NeT, below 100 PeV we used the one reported in Ref.~\cite{KM3NeT_poster} (purple solid), while above 100 PeV we extrapolate it by matching our own effective area calculation (purple dashed, see text for details). 
    We take the all-sky averaged KM3NeT effective area (gold) from Ref.~\cite{KM3NeT:2025npi}. 
    In the lower panel we plot the ratio of effective areas for  IceCube $8^\circ$ declination and KM3NeT all sky. 
    } 
    \label{fig:Aeff}
\end{figure}

Figure~\ref{fig:Aeff} shows the $A_\text{eff}$ used throughout this work. 
IceCube effective area at $8^\circ$ above the horizon (green) comes from an IceCube point source search~\cite{IceCube:2021xar}. 
{Note that this effective area accounts for background rejection and therefore has a higher threshold than in the background free search at the horizon.}
Below $100$ PeV, KM3NeT data at horizon (purple) can be found in Ref.~\cite{KM3NeT_poster} while at higher energies we use our own  calculation of the effective area, matching the normalization to account for the simplifications made here. 
We find the energy dependence of $A_{\rm eff}$ to match the KM3NeT curve well and, based on our calculation, we can extend the effective area to higher energies, and arbitrary zenith angles.
For the all-sky averaged $A_\text{eff}$ (gold) we use the result provided by the KM3NeT collaboration~\cite{KM3NeT:2025npi}. 

\section{Neutrino and Muon Energy Reconstruction}\label{app:Nhit}

\begin{figure}[t]
    \centering
    \includegraphics[width=1.0\linewidth]{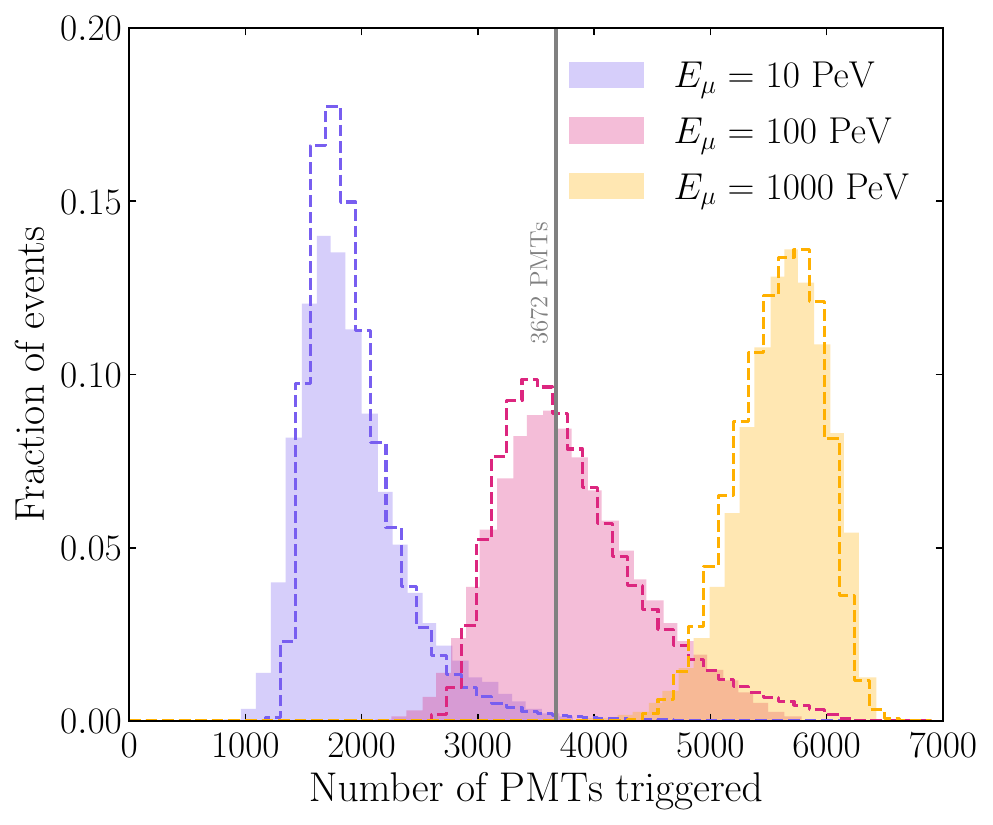}
    \caption{The distributions derived in this work of the number of PMTs triggered for 10, 100, and 1000 PeV muons pass through KM3NeT as dashed histograms. The official results are also shown for comparison as shaded histograms~\cite{KM3NeT:2025npi}.
    The vertical line labeled ``3,672 PMTs'' indicates the number of triggered PMTs for the high energy event we are interested in.}
    \label{fig:validation}
\end{figure}

As discussed in the main text, Bayes theorem states that the probability of a neutrino energy given the number of triggered PMTs (hits) is
\begin{equation}
    P(E_\nu|N_{\rm hit}) = \frac{1}{P(N_{\rm hit})} \int dE_\mu P(N_{\rm hit}|E_\mu)P(E_\mu|E_\nu)P(E_\nu),
\end{equation}
where $P(E_\nu)$ is the prior on the neutrino energy, namely
\begin{equation}
    P(E_\nu) = \frac{1}{\phi_\nu}\frac{d\phi_\nu}{dE_\nu},
\end{equation}
where $\phi_\nu$ is the neutrino flux.
$P(E_\mu|E_\nu)$ is the probability that a neutrino with energy $E_\nu$ interacts and produces a muon which has energy $E_\mu$ when entering the detector.
This probability accounts for muon production from a neutrino with given energy, which is simulated with \texttt{MadGraph}~\cite{Alwall:2014hca}.

More concretely,
\begin{align}
    P(E_\mu|E_\nu) = \frac{1}{\sigma(E_\nu)}
    \frac{d\tilde \sigma(E_\nu,E_\mu)}{dE_\mu},
\end{align}
where $\sigma(E_\nu)$ is the neutrino cross section. 
The differential cross section for a neutrino with energy $E_\nu$ to produce a muon which reaches the detector with energy $E_\mu$ is given by~\cite{Gaisser:2016uoy}
\begin{equation}
    \frac{d\tilde\sigma}{dE_\mu}  = 
      \frac{N_A}{\alpha(1 + E_\mu/E_{\rm crit})} \int_0^{1-E_\mu/E_\nu}\hspace{-2mm} dy' \int_0^1 dx \frac{d\sigma}{dx dy'},
\end{equation}
where $\alpha\simeq 2$~MeV/cm$^2$; $E_{\rm crit}$ is the critical energy, typically around 500 GeV~\cite{Gaisser:2016uoy}; $x$ is the Bjorken-$x$; and $y'$ is the inelasticity parameter.

We then calculate the probability $P(N_{\rm hit}|E_\mu)$ that a muon arriving at the detector with energy $E_\mu$ triggers a certain number of PMTs $N_{\rm hit}$.
We simulate muon energy loss in the detector with \texttt{PROPOSAL}~\cite{koehne2013proposal,dunsch_2020_1484180,dunsch_2018_proposal_improvements} which records muon energy deposition through both continuous and radiative energy losses. 
We then map the muon energy loss to a number of triggered PMTs, $N_{\rm hit}$, in a simplified manner. 
We assume that light will be attenuated exponentially as a function of propagation distance in water. 
A PMT will trigger if the light intensity reaching it is above some threshold, which we take to be 8 GeV/cm$^2$.
Varying this threshold does not lead to a large variation on $N_{\rm hit}$.
This allows us to compute the fraction of detector volume which is above threshold and convert it to a number of triggered PMTs assuming they are uniformly distributed through the detector volume.

We find optimal values for light attenuation and PMT thresholds by matching the KM3NeT distributions for 10, 100, and 1000 PeV muons as shown in Fig.~\ref{fig:validation}.
To model the tail of the $N_{\rm hit}$ distribution, we fit our histograms to a Fr\'echet distribution
\begin{equation}\label{eq:frechet}
    P(x) = \frac{\alpha}{s}\left(\frac{x}{s}\right)^{-1-\alpha}e^{-\left(x/s\right)^{-\alpha}},
\end{equation}
where $\alpha$ sets the width and $s$ the peak location. 
In Fig.~\ref{fig:validation} we compare our distributions (dashed histograms) to KM3NeT's official ones (shaded histograms).
For reference, the uncertainties on light propagation in water shift the KM3NeT distribution peaks by about 10\%~\cite{KM3NeT_pres}. 
For the highest energy muons, we simulate the truncation of the triggered PMTs around 6,000 with a hyperbolic tangent function to match the KM3NeT result.

\bibliography{KM3NeT-IceCube}

@article{Baikal-GVD:2018isr,
    author = "Avrorin, A. D. and others",
    editor = "Volkova, V. E. and Zhezher, Y. V. and Levkov, D. G. and Rubakov, V. A. and Matveev, V. A.",
    collaboration = "Baikal-GVD",
    title = "{Baikal-GVD: status and prospects}",
    eprint = "1808.10353",
    archivePrefix = "arXiv",
    primaryClass = "astro-ph.IM",
    doi = "10.1051/epjconf/201819101006",
    journal = "EPJ Web Conf.",
    volume = "191",
    pages = "01006",
    year = "2018"
}

@article{P-ONE:2020ljt,
    author = "Agostini, Matteo and others",
    collaboration = "P-ONE",
    title = "{The Pacific Ocean Neutrino Experiment}",
    eprint = "2005.09493",
    archivePrefix = "arXiv",
    primaryClass = "astro-ph.HE",
    doi = "10.1038/s41550-020-1182-4",
    journal = "Nature Astron.",
    volume = "4",
    number = "10",
    pages = "913--915",
    year = "2020"
}

@article{IceCube-Gen2:2020qha,
    author = "Aartsen, M. G. and others",
    collaboration = "IceCube-Gen2",
    title = "{IceCube-Gen2: the window to the extreme Universe}",
    eprint = "2008.04323",
    archivePrefix = "arXiv",
    primaryClass = "astro-ph.HE",
    doi = "10.1088/1361-6471/abbd48",
    journal = "J. Phys. G",
    volume = "48",
    number = "6",
    pages = "060501",
    year = "2021"
}

@article{KM3NeT:2025npi,
    author = "Aiello, S. and others",
    collaboration = "KM3NeT",
    title = "{Observation of an ultra-high-energy cosmic neutrino with KM3NeT}",
    doi = "10.1038/s41586-024-08543-1",
    journal = "Nature",
    volume = "638",
    number = "8050",
    pages = "376--382",
    year = "2025"
}

@misc{KM3NeT:2025ccp,
    author = "Adriani, O. and others",
    collaboration = "KM3NeT",
    title = "{The ultra-high-energy event KM3-230213A within the global neutrino landscape}",
    eprint = "2502.08173",
    archivePrefix = "arXiv",
    primaryClass = "astro-ph.HE",
    month = "2",
    year = "2025"
}

@article{Zatsepin:1966jv,
    author = "Zatsepin, G. T. and Kuzmin, V. A.",
    title = "{Upper limit of the spectrum of cosmic rays}",
    journal = "JETP Lett.",
    volume = "4",
    pages = "78--80",
    year = "1966"
}

@article{vanVliet:2019nse,
    author = {van Vliet, Arjen and Alves Batista, Rafael and H\"orandel, J\"org R.},
    title = "{Determining the fraction of cosmic-ray protons at ultrahigh energies with cosmogenic neutrinos}",
    eprint = "1901.01899",
    archivePrefix = "arXiv",
    primaryClass = "astro-ph.HE",
    reportNumber = "DESY-18-215",
    doi = "10.1103/PhysRevD.100.021302",
    journal = "Phys. Rev. D",
    volume = "100",
    number = "2",
    pages = "021302",
    year = "2019"
}

@article{Ahlers:2010fw,
    author = "Ahlers, M. and Anchordoqui, L. A. and Gonzalez-Garcia, M. C. and Halzen, F. and Sarkar, S.",
    title = "{GZK Neutrinos after the Fermi-LAT Diffuse Photon Flux Measurement}",
    eprint = "1005.2620",
    archivePrefix = "arXiv",
    primaryClass = "astro-ph.HE",
    reportNumber = "YITP-SB-10-014",
    doi = "10.1016/j.astropartphys.2010.06.003",
    journal = "Astropart. Phys.",
    volume = "34",
    pages = "106--115",
    year = "2010"
}

@article{Ahlers:2012rz,
    author = "Ahlers, Markus and Halzen, Francis",
    title = "{Minimal Cosmogenic Neutrinos}",
    eprint = "1208.4181",
    archivePrefix = "arXiv",
    primaryClass = "astro-ph.HE",
    doi = "10.1103/PhysRevD.86.083010",
    journal = "Phys. Rev. D",
    volume = "86",
    pages = "083010",
    year = "2012"
}

@inproceedings{Meier:2024flg,
    author = "Meier, Maximilian",
    collaboration = "IceCube",
    title = "{Recent cosmogenic neutrino search results with IceCube and prospects with IceCube-Gen2}",
    booktitle = "{58th Rencontres de Moriond on Very High Energy Phenomena in the Universe}",
    eprint = "2409.01740",
    archivePrefix = "arXiv",
    primaryClass = "astro-ph.HE",
    month = "9",
    year = "2024"
}

@article{Greisen:1966jv,
    author = "Greisen, Kenneth",
    title = "{End to the cosmic ray spectrum?}",
    doi = "10.1103/PhysRevLett.16.748",
    journal = "Phys. Rev. Lett.",
    volume = "16",
    pages = "748--750",
    year = "1966"
}

@article{Silva:2023wol,
    author = "Silva, Manuel and Mancina, Sarah and Osborn, Jesse",
    collaboration = "IceCube",
    title = "{Measurement of the Cosmic Neutrino Flux from the Southern Sky using 10 years of IceCube Starting Track Events}",
    eprint = "2308.04582",
    archivePrefix = "arXiv",
    primaryClass = "astro-ph.HE",
    reportNumber = "PoS-ICRC2023-1008",
    doi = "10.22323/1.444.1008",
    journal = "PoS",
    volume = "ICRC2023",
    pages = "1008",
    year = "2023"
}

@article{Alwall:2014hca,
    author = "Alwall, J. and Frederix, R. and Frixione, S. and Hirschi, V. and Maltoni, F. and Mattelaer, O. and Shao, H. -S. and Stelzer, T. and Torrielli, P. and Zaro, M.",
    title = "{The automated computation of tree-level and next-to-leading order differential cross sections, and their matching to parton shower simulations}",
    eprint = "1405.0301",
    archivePrefix = "arXiv",
    primaryClass = "hep-ph",
    reportNumber = "CERN-PH-TH-2014-064, CP3-14-18, LPN14-066, MCNET-14-09, ZU-TH-14-14",
    doi = "10.1007/JHEP07(2014)079",
    journal = "JHEP",
    volume = "07",
    pages = "079",
    year = "2014"
}

@misc{IceCube:2021xar,
    author = "Abbasi, R. and others",
    collaboration = "IceCube",
    title = "{IceCube Data for Neutrino Point-Source Searches Years 2008-2018}",
    eprint = "2101.09836",
    archivePrefix = "arXiv",
    primaryClass = "astro-ph.HE",
    doi = "10.21234/CPKQ-K003",
    month = "1",
    year = "2021"
}

@article{IceCube:2020wum,
    author = "Abbasi, R. and others",
    collaboration = "IceCube",
    title = "{The IceCube high-energy starting event sample: Description and flux characterization with 7.5 years of data}",
    eprint = " ",
    archivePrefix = "arXiv",
    primaryClass = "astro-ph.HE",
    doi = "10.1103/PhysRevD.104.022002",
    journal = "Phys. Rev. D",
    volume = "104",
    pages = "022002",
    year = "2021"
}

@unpublished{KM3Net_poster,
    author = "Caiffi, B. and others",
    collaboration = "KM3Net",
    title = "Combined diffuse and point source search with KM3NeT/ARCA and ANTARES neutrino telescopes",
    year = "2024",
    note = "Poster"
}

@article{IceCube:2021uhz,
    author = "Abbasi, R. and others",
    collaboration = "IceCube",
    title = "{Improved Characterization of the Astrophysical Muon\textendash{}neutrino Flux with 9.5 Years of IceCube Data}",
    eprint = "2111.10299",
    archivePrefix = "arXiv",
    primaryClass = "astro-ph.HE",
    doi = "10.3847/1538-4357/ac4d29",
    journal = "Astrophys. J.",
    volume = "928",
    number = "1",
    pages = "50",
    year = "2022"
}

@article{IceCube:2020acn,
    author = "Aartsen, M. G. and others",
    collaboration = "IceCube",
    title = "{Characteristics of the diffuse astrophysical electron and tau neutrino flux with six years of IceCube high energy cascade data}",
    eprint = "2001.09520",
    archivePrefix = "arXiv",
    primaryClass = "astro-ph.HE",
    doi = "10.1103/PhysRevLett.125.121104",
    journal = "Phys. Rev. Lett.",
    volume = "125",
    number = "12",
    pages = "121104",
    year = "2020"
}

@inproceedings{Naab:2023xcz,
    author = "Naab, Richard and Ganster, Erik and Zhang, Zelong",
    collaboration = "IceCube",
    title = "{Measurement of the astrophysical diffuse neutrino flux in a combined fit of IceCube's high energy neutrino data}",
    booktitle = "{38th International Cosmic Ray Conference}",
    eprint = "2308.00191",
    archivePrefix = "arXiv",
    primaryClass = "astro-ph.HE",
    reportNumber = "PoS-ICRC2023-1064",
    month = "7",
    year = "2023"
}

@misc{Aiello:2024jbp,
    author = "Aiello, S. and others",
    title = "{Astronomy potential of KM3NeT/ARCA}",
    eprint = "2402.08363",
    archivePrefix = "arXiv",
    primaryClass = "astro-ph.HE",
    month = "2",
    year = "2024"
}

@conference{KM3Net_pres,
    author = "Coelho, J",
    booktitle = "XXXI International Conference on Neutrino Physics and Astrophysics",
    title = "Latest results from KM3NeT",
    year = "2024",
    month = "6",
    note = "on behalf of the KM3NeT Collaboration"
}

@article{IceCube:2013cdw,
    author = "Aartsen, M. G. and others",
    collaboration = "IceCube",
    title = "{First observation of PeV-energy neutrinos with IceCube}",
    eprint = "1304.5356",
    archivePrefix = "arXiv",
    primaryClass = "astro-ph.HE",
    doi = "10.1103/PhysRevLett.111.021103",
    journal = "Phys. Rev. Lett.",
    volume = "111",
    pages = "021103",
    year = "2013"
}

@article{IceCube:2013low,
    author = "Aartsen, M. G. and others",
    collaboration = "IceCube",
    title = "{Evidence for High-Energy Extraterrestrial Neutrinos at the IceCube Detector}",
    eprint = "1311.5238",
    archivePrefix = "arXiv",
    primaryClass = "astro-ph.HE",
    doi = "10.1126/science.1242856",
    journal = "Science",
    volume = "342",
    pages = "1242856",
    year = "2013"
}

@article{IceCube:2022der,
    author = "Abbasi, R. and others",
    collaboration = "IceCube",
    title = "{Evidence for neutrino emission from the nearby active galaxy NGC 1068}",
    eprint = "2211.09972",
    archivePrefix = "arXiv",
    primaryClass = "astro-ph.HE",
    doi = "10.1126/science.abg3395",
    journal = "Science",
    volume = "378",
    number = "6619",
    pages = "538--543",
    year = "2022"
}

@article{IceCube:2024fxo,
    author = "Abbasi, R. and others",
    collaboration = "IceCube",
    title = "{Characterization of the astrophysical diffuse neutrino flux using starting track events in IceCube}",
    eprint = "2402.18026",
    archivePrefix = "arXiv",
    primaryClass = "astro-ph.HE",
    doi = "10.1103/PhysRevD.110.022001",
    journal = "Phys. Rev. D",
    volume = "110",
    number = "2",
    pages = "022001",
    year = "2024"
}

@article{KM3Net:2016zxf,
    author = "Adrian-Martinez, S. and others",
    collaboration = "KM3Net",
    title = "{Letter of intent for KM3NeT 2.0}",
    eprint = "1601.07459",
    archivePrefix = "arXiv",
    primaryClass = "astro-ph.IM",
    doi = "10.1088/0954-3899/43/8/084001",
    journal = "J. Phys. G",
    volume = "43",
    number = "8",
    pages = "084001",
    year = "2016"
}

@book{Gaisser:2016uoy,
    author = "Gaisser, Thomas K. and Engel, Ralph and Resconi, Elisa",
    title = "{Cosmic Rays and Particle Physics}: {2nd Edition}",
    isbn = "978-0-521-01646-9",
    publisher = "Cambridge University Press",
    month = "6",
    year = "2016"
}

@article{Dziewonski:1981xy,
    author = "Dziewonski, A. M. and Anderson, D. L.",
    title = "{Preliminary reference earth model}",
    doi = "10.1016/0031-9201(81)90046-7",
    journal = "Phys. Earth Planet. Interiors",
    volume = "25",
    pages = "297--356",
    year = "1981"
}

@article{koehne2013proposal,
  title     ={PROPOSAL: A tool for propagation of charged leptons},
  author    = {Koehne, Jan-Hendrik and
               Frantzen, Katharina and
               Schmitz, Martin and
               Fuchs, Tomasz and
               Rhode, Wolfgang and
               Chirkin, Dmitry and
               Tjus, J Becker},
  journal   = {Computer Physics Communications},
  volume    = {184},
  number    = {9},
  pages     = {2070--2090},
  year      = {2013},
  doi       = {10.1016/j.cpc.2013.04.001}
}

@misc{dunsch_2020_1484180,
  author = {Alameddine, Jean-Marco and
            Dunsch, Mario and
            Bollmann, Lars and
            Fuchs, Tomasz and
            Gutjahr, Pascal and
            Koehne, Jan-Hendrik and
            Kopper, Claudio and
            Krings, Kai and
            Kuo, Chung-Yun and
            Menne, Thorben and
            Noethe, Maximilian and
            Olivas, Alex and
            Rhode, Wolfgang and
            Sackel, Maximilian and
            Sandrock, Alexander and
            Schneider, Austin and
            Soedingrekso, Jan and
            van Santen, Jacob},
  title  = {tudo-astroparticlephysics/PROPOSAL: Zenodo},
  month  = mar,
  year   = 2020,
  doi    = {10.5281/zenodo.1484180},
  url    = {https://doi.org/10.5281/zenodo.1484180}
}

@article{dunsch_2018_proposal_improvements,
  title       = {Recent Improvements for the Lepton Propagator PROPOSAL},
  author      = {Dunsch, Mario and
                 Soedingrekso, Jan and
                 Sandrock, Alexander and
                 Meier, Max and
                 Menne, Thorben and
                 Rhode, Wolfgang},
  journal     = {Computer Physics Communications},
  volume      = {242},
  pages       = {132--144},
  year        = {2019},
  eprint      = {1809.07740},
  eprinttype  = {arXiv},
  eprintclass = {hep-ph},
  doi         = {10.1016/j.cpc.2019.03.021}
}

@misc{IceCube:2025ezc,
    author = "Abbasi, R. and others",
    collaboration = "IceCube",
    title = "{A search for extremely-high-energy neutrinos and first constraints on the ultra-high-energy cosmic-ray proton fraction with IceCube}",
    eprint = "2502.01963",
    archivePrefix = "arXiv",
    primaryClass = "astro-ph.HE",
    month = "2",
    year = "2025"
}

@misc{neronov2025,
      title={KM3-230213A: An Ultra-High Energy Neutrino from a Year-Long Astrophysical Transient}, 
      author={Andrii Neronov and Foteini Oikonomou and Dmitri Semikoz},
      year={2025},
      eprint={2502.12986},
      archivePrefix={arXiv},
      primaryClass={astro-ph.HE},
      url={https://arxiv.org/abs/2502.12986}, 
}

@misc{KM3NeT:2025aps,
    author = "Adriani, O. and others",
    collaboration = "KM3NeT",
    title = "{On the Potential Galactic Origin of the Ultra-High-Energy Event KM3-230213A}",
    eprint = "2502.08387",
    archivePrefix = "arXiv",
    primaryClass = "astro-ph.HE",
    month = "2",
    year = "2025"
}

@misc{KM3NeT:2025bxl,
    author = "Adriani, O. and others",
    collaboration = "KM3NeT, MessMapp Group, Fermi-LAT, Owens Valley Radio Observatory 40-m Telescope Group, SVOM",
    title = "{Characterising Candidate Blazar Counterparts of the Ultra-High-Energy Event KM3-230213A}",
    eprint = "2502.08484",
    archivePrefix = "arXiv",
    primaryClass = "astro-ph.HE",
    month = "2",
    year = "2025"
}

@misc{Muzio:2025gbr,
    author = "Muzio, Marco S. and Yuan, Tianlu and Lu, Lu",
    title = "{Emergence of a neutrino flux above 5 PeV and implications for ultrahigh energy cosmic rays}",
    eprint = "2502.06944",
    archivePrefix = "arXiv",
    primaryClass = "astro-ph.HE",
    month = "2",
    year = "2025"
}

@misc{Fang:2025nzg,
    author = "Fang, Ke and Halzen, Francis and Hooper, Dan",
    title = "{Cascaded Gamma-ray Emission Associated with the KM3NeT Ultra-High-Energy Event KM3-230213A}",
    eprint = "2502.09545",
    archivePrefix = "arXiv",
    primaryClass = "astro-ph.HE",
    month = "2",
    year = "2025"
}

@misc{Dzhatdoev:2025sdi,
    author = "Dzhatdoev, T. A.",
    title = "{The blazar PKS 0605-085 as the origin of the KM3-230213A ultra high energy neutrino event}",
    eprint = "2502.11434",
    archivePrefix = "arXiv",
    primaryClass = "astro-ph.HE",
    month = "2",
    year = "2025"
}
\end{document}